\documentclass[multphys,vecphys]{svmult}

\usepackage{makeidx}         
\usepackage{graphicx}        
\usepackage{multicol}        
\usepackage[bottom]{footmisc}

\makeindex             

\begin{document}

\title*{Tidal tails around globular clusters: are they good tracers of cluster orbits?}
\titlerunning{Tidal tails around globular clusters} 
\author{P. Di Matteo\inst{1,2}, R. Capuzzo Dolcetta\inst{2},
P. Miocchi\inst{3,2} \and M. Montuori\inst{4,2}}
\institute{Obs. de Paris, 61, Av. de L'Observatoire
75014 Paris, France 
\texttt{paola.dimatteo@obspm.fr}
\and Dip. di Fisica, Universit\'a di Roma La Sapienza, P.le Aldo Moro 2, 00185 Roma, Italia  
\texttt{roberto.capuzzodolcetta@uniroma1.it} 
\and INAF - Oss. di Teramo, via M. Maggini, 64100 Teramo, Italia \texttt{miocchi@uniroma1.it}
\and CNR - Ist. Sistemi Complessi, Roma, Italia \texttt{montuorm@roma1.infn.it}}
%
%
\maketitle
\section*{Introduction}

In the last decade, observational studies have shown the existence of tidal streams in the outer part of many galactic globular clusters \cite{grill,leon}. The most striking examples of clusters with well defined tidal tails are represented by Palomar 5 \cite{oden} and NGC 5466 \cite{bel} (both observed in the framework of the Sloan Digital Sky Survey), which show structures elongated for 4 kpc and 1 kpc in length, respectively.  Unfortunately, most of the observational studies about globular clusters (GCs) do not cover such a large field of the sky as the SDSS does. \\
In this framework, by mean of a parallel, adaptive tree-code \cite{miocchi}, we performed detailed N-body simulations of GCs moving in a realistic three-components (bulge, disk and halo) Milky Way potential \cite{allen}, in order to clarify  whether and to what extent tails in the clusters outer regions (few tidal radii) are tracers of the local orbits and, also, if some kind of correlation exists among the cluster orbital phase and the orientation of such streams. 
\section*{Results}
\begin{figure}
\centering
\includegraphics[height=14cm,angle=270]{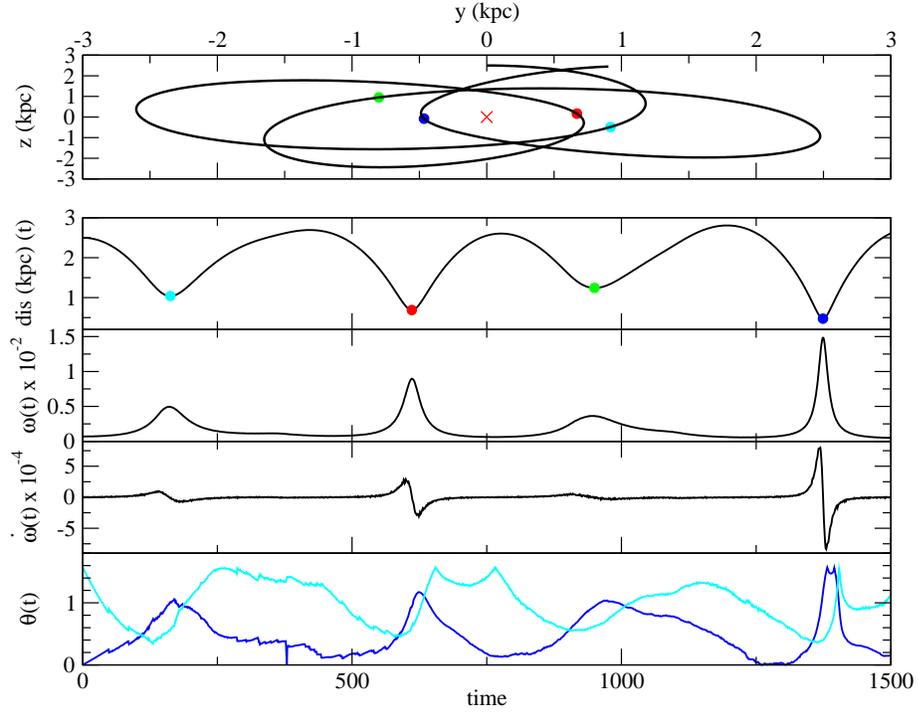}
\caption{From top to the bottom: 1.Plot of the GC orbits; 2. Distance of the GC from the Galaxy center as a function of time; 3. $|\mbox{\boldmath{$\omega$}}|$, as a function of time; 4. Derivative with time of  $| \mbox{\boldmath{$\omega$}}|$; 5. Angles formed by the inner part of the tails with the galactic center direction (blue curve) and with the GC local orbit (cyan curve).
A strong correlation exists between tails alignment and GC orbital angular velocity.
}
\label{fig2}       
\end{figure}

While the outer part of the tails are good tracers of GC trajectory, the inner part is never aligned with the GC path, except for more eccentric orbits, when the GC moves towards the pericenter \cite{cdm05,dcm}. 
For a given orbit, a strong correlation exists between tails alignment and GC orbital angular velocity (see  Fig.\ref{fig2}): when  $| \mbox{\boldmath{$\omega$}}|$ increases (moving to pericenter), tails are stretched along the GC path (the angle between tails and GC orbital velocity decreases and reaches its minimum value just before the pericenter passage); when  $| \mbox{\boldmath{$\omega$}}|$ decreases (moving to apocenter) tails deviate from the GC path and, in turn, the angle between tails and galactic center (blue curve in Fig.\ref{fig2}) diminishes, indicating that, in this phase, tails are more radially pointing.  
For a discussion, see also \cite{cdm05,dcm,prep}.\\

%
%

%
%



\printindex
\end{document}